\documentclass{llncs}
\usepackage{rotating}
\usepackage[final]{microtype}
\usepackage{color}
\usepackage{latexsym}
\usepackage{amssymb}
\usepackage{amsmath}
\usepackage{xspace}
\usepackage{multirow}
\usepackage{wrapfig}
\usepackage{url}
\usepackage{bussproofs}
\usepackage[hypertexnames=false]{hyperref}
\usepackage[caption=false, position=top]{subfig}

\newcommand{\quaffle}{\textsf{Quaffle}\xspace}
\newcommand{\bloqqer}{\textsf{Bloqqer}\xspace}
\newcommand{\aigsolve}{\textsf{AIGSolve}\xspace}
\newcommand{\ghostqcegar}{\textsf{GhostQ}\xspace}

\newcommand{\qstsbreaksym}{\textsf{QSTS}\xspace}

\newcommand{\depqbfbat}{\textsf{DQ}\xspace}
\newcommand{\depqbfnoclauseaxioms}{\textsf{DQ-ncl}\xspace}
\newcommand{\depqbfnocubeaxioms}{\textsf{DQ-ncu}\xspace}
\newcommand{\depqbfnoaxioms}{\textsf{DQ-n}\xspace}
\newcommand{\depqbfnoqbce}{\textsf{DQ-nq}\xspace}

\newcommand{\depqbfdepmansimple}{\textsf{DQ-lin}\xspace}
\newcommand{\depqbfdepmansimpleldq}{\textsf{DQ-linldq}\xspace}

\newcommand{\caqe}{\textsf{CAQE}\xspace}
\newcommand{\qesto}{\textsf{QESTO}\xspace}
\newcommand{\rareqs}{\textsf{RAReQS}\xspace}

\newcommand{\depqbf}{\textsf{DepQBF}\xspace}
\newcommand{\depqbfnewversion}{\textsf{DepQBF~6.0}\xspace}
\newcommand{\depqbfnewversioncurrent}{\textsf{DepQBF~6.02}\xspace}
\newcommand{\nenofex}{\textsf{Nenofex}\xspace}

\newcommand{\UR}{\mathit{UR}}
\newcommand{\ER}{\mathit{ER}}

\newcommand{\lit}[1]{\mathsf{var}(#1)}
\newcommand{\prefix}{\Pi}
\newcommand{\clauset}{\phi}
\newcommand{\qclauset}{\psi}
\newcommand{\quant}[2]{\mathsf{Q}(#1,#2)}
\newcommand{\satequiv}{\equiv_{\mathit{sat}}}
\newcommand{\qrescalc}{QRES\xspace}
\newcommand{\LCL}{\theta}
\newcommand{\LCU}{\gamma}

\author{Florian Lonsing \and Uwe Egly}

\institute{Knowledge-Based Systems Group, 
  Vienna University of Technology, Austria
  \\ \email{firstname.lastname@tuwien.ac.at}
}

\begin{document}

\newcommand{\doctitle}{\depqbfnewversion: A Search-Based QBF Solver Beyond Traditional QCDCL}

\title{\doctitle
\thanks{Supported by the Austrian Science Fund (FWF) under grant
S11409-N23. This article will appear in the \textbf{proceedings} of the
\emph{26th International Conference on Automated Deduction (CADE-26)}, LNCS,
Springer, 2017.}}

\maketitle

\begin{abstract}
We present the latest major release 
version~6.0 of the \emph{quantified Boolean formula (QBF)} solver \depqbf, which is
based on \emph{QCDCL}. QCDCL is an extension of the \emph{conflict-driven
clause learning (CDCL)} paradigm implemented in state of the art propositional
satisfiability (SAT) solvers.  The \emph{Q-resolution calculus (\qrescalc)} is
a QBF proof system which underlies QCDCL.  QCDCL solvers can produce \qrescalc
proofs of QBFs in \emph{prenex conjunctive normal form (PCNF)} as a byproduct of the
solving process.  In contrast to traditional QCDCL based on \qrescalc,
\depqbfnewversion implements a variant of QCDCL which is based on a
generalization of \qrescalc. This generalization is due to a set of additional
axioms and leaves the original Q-resolution rules unchanged. The
generalization of \qrescalc enables QCDCL to potentially produce exponentially
shorter proofs than the traditional variant.  We present an overview of the features
implemented in \depqbf and report on experimental results which demonstrate
the effectiveness of generalized \qrescalc in QCDCL.
\end{abstract}


\section{Introduction}
\label{sec:intro}

Propositional satisfiability (SAT) solvers based on \emph{conflict-driven
clause learning
(CDCL)}~\cite{DBLP:series/faia/SilvaLM09}
implement 
a combination of the DPLL algorithm~\cite{DBLP:journals/cacm/DavisLL62} and
propositional \emph{resolution}~\cite{DBLP:journals/jacm/Robinson65} to derive
\emph{learned clauses} from a CNF to be solved.

CDCL has been extended to solve \emph{quantified Boolean formulas
(QBFs)}~\cite{DBLP:series/faia/BuningB09}, resulting in the \emph{QCDCL}
approach~\cite{DBLP:journals/jair/GiunchigliaNT06,DBLP:conf/tableaux/Letz02,DBLP:conf/iccad/ZhangM02}. The
logic of QBFs allows for explicit universal and existential quantification of
propositional variables. As a consequence, the satisfiability problem of QBFs
is PSPACE-complete.

In contrast to SAT solving, where CDCL is the dominant solving paradigm in
practice, QCDCL is complemented by \emph{variable
expansion}~\cite{DBLP:conf/fmcad/AyariB02,DBLP:conf/sat/Biere04a}. This
approach successively eliminates variables from a QBF until it reduces to
either true or false. Many modern solvers
(e.g.~\cite{Janota20161,DBLP:conf/ijcai/JanotaM15,DBLP:conf/fmcad/RabeT15})
implement expansion by \emph{counter-example guided abstraction refinement
(CEGAR)}~\cite{DBLP:journals/jacm/ClarkeGJLV03}.

The \emph{Q-resolution calculus
(\qrescalc)}~\cite{DBLP:journals/jair/GiunchigliaNT06,DBLP:journals/iandc/BuningKF95,DBLP:conf/tableaux/Letz02,DBLP:conf/iccad/ZhangM02}
is a QBF proof system that underlies QCDCL in a way that is analogous
to propositional resolution in CDCL. The empty clause is derivable
from a PCNF $\qclauset$ by \qrescalc iff $\qclauset$ is
unsatisfiable. According to QBF proof complexity, there is an
exponential separation between the sizes of proofs that variable expansion
and Q-resolution can
produce for certain QBFs~\cite{beyersdorff_et_al:LIPIcs:2015:4905,DBLP:journals/tcs/JanotaM15}.
This
theoretical result suggests to combine such orthogonal proof systems in QBF
solvers to leverage their individual strengths.

As a first step towards a solver framework that allows for the combination of QBF
proof systems in a systematic way, we present the latest major release version~6.0 of the QCDCL solver
\depqbf.\footnote{\depqbf is licensed under GPLv3:
\url{http://lonsing.github.io/depqbf/}} In contrast to traditional
QCDCL based on
\qrescalc~\cite{DBLP:journals/jair/GiunchigliaNT06,DBLP:journals/iandc/BuningKF95,DBLP:conf/tableaux/Letz02,DBLP:conf/iccad/ZhangM02},
\depqbfnewversion implements a variant of QCDCL that relies on a generalization of
\qrescalc. This generalization is due to a set of new axioms added to
\qrescalc~\cite{DBLP:conf/sat/LonsingES16}. In practice, derivations made
by the added axioms in QCDCL are based on \emph{arbitrary} QBF proof
systems. As a consequence, when applying proof systems that are orthogonal to
Q-resolution, the generalization of \qrescalc via the new axioms enables QCDCL as
implemented in \depqbfnewversion to potentially produce exponentially shorter
proofs than traditional QCDCL. 

We report on experiments where we compare
\depqbfnewversion to state of the art QBF solvers. Experimental results
demonstrate the effectiveness of generalized \qrescalc in QCDCL. 
Additionally, we briefly summarize the evolution of \depqbf since the first 
version~0.1~\cite{DBLP:journals/jsat/LonsingB10}. We relate the features that
were added to the different versions of \depqbf over time to the enhanced variant of QCDCL
implemented in \depqbfnewversion.


\section{Preliminaries}
\label{sec:prelims}

A QBF $\qclauset := \prefix.\clauset$ in \emph{prenex conjunctive normal form (PCNF)} 
consists of a \emph{quantifier prefix} $\prefix := Q_1X_1 \ldots Q_nX_n$ and a
CNF $\clauset$ not containing tautological clauses. 
The CNF $\clauset$ is defined over the propositional variables $X_1 \cup \ldots \cup X_n$  
that appear in $\prefix$. The variable sets $X_i$ are pairwise disjoint and $Q_i
\not= Q_{i+1}$ for 
$Q_i \in \{\forall, \exists\}$. 
QBFs $\qclauset := \prefix.\clauset$ in \emph{prenex disjunctive normal
  form (PDNF)} are defined analogously to PCNFs, where $\clauset$ is a DNF
consisting of \emph{cubes}. A cube is a conjunction of literals.
 The \emph{quantifier}
$\quant{\prefix}{l}$ of  a literal $l$ is  $Q_i$ if the variable $\lit{l}$ of $l$ appears in
$X_i$. If $\quant{\prefix}{l} = Q_i$
and $\quant{\prefix}{k} = Q_j$,
then \nolinebreak $l \leq_\prefix k$ \nolinebreak iff \nolinebreak $i \leq j$.

An \emph{assignment} $A$ maps variables  
of a QBF $\prefix.\clauset$ to truth values \emph{true} ($\top$) and
\emph{false} ($\bot$). We
represent $A = \{l_1,\ldots,l_n\}$ 
as a set of literals such that if a variable $x$ is
assigned \emph{true} (\emph{false}) then $l_i \in A$ with $l_i = x$ ($l_i =
\bar x$), where $\bar x$ is the negation of $x$.  
Further, $\lit{l_i} \not= \lit{l_j}$ for any $l_i, l_j \in A$ with $i \not= j$. 

The PCNF $\qclauset$ \emph{under assignment} $A$, 
written as 
$\qclauset[A]$, is the PCNF
obtained from $\qclauset$ in which 
for all $l \in A$, all clauses containing $l$ are removed, 
all occurrences of $\bar l$ are deleted, and $\lit{l}$ is 
removed from the prefix. 
If the CNF of $\qclauset[A]$ is empty (respectively, contains the empty clause $\emptyset$), then 
it is satisfied (falsified) by $A$ and 
$A$ is a \emph{satisfying} (\emph{falsifying}) assignment, written as $\qclauset[A] = \top$ ($\qclauset[A] = \bot$).
A \emph{PDNF $\qclauset$ under an assignment
$A$} and an \emph{empty cube} are defined in a way dual to PCNFs and empty clauses. 
A QBF $\prefix.\clauset$ 
with $Q_1 = \exists$ ($Q_1 = \forall$) 
is satisfiable iff, for $x \in X_1$, 
$\prefix.\clauset[\{x\}]$ or (and)
$\prefix.\clauset[\{\bar x\}]$ is satisfiable. 
Two QBFs $\qclauset$ and $\qclauset'$ are \emph{satisfiability-equivalent} 
($\qclauset \satequiv \qclauset'$), iff
$\qclauset$ is satisfiable whenever $\qclauset'$ is
satisfiable. 


\section{QCDCL and the Generalized Q-Resolution Calculus}
\label{sec:qcdcl}

\begin{figure}[t]
\centering
\includegraphics{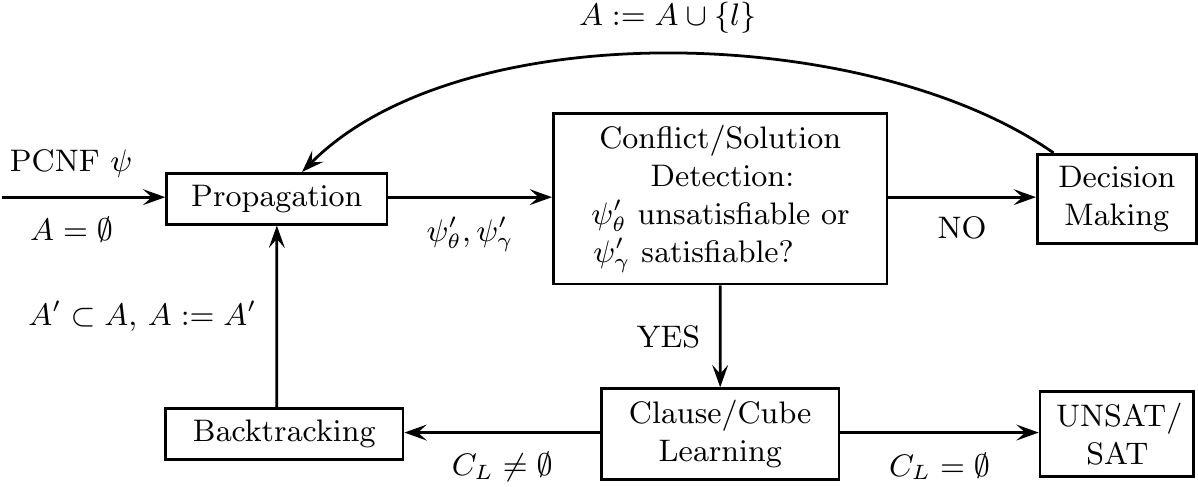}
\caption{Workflow of the variant of QCDCL implemented in \depqbfnewversion
that relies on a generalization of the Q-resolution calculus (\qrescalc)
(figure adapted from~\cite{DBLP:conf/sat/LonsingES16}).}
\label{fig:qcdcl:enhanced}
\end{figure}

In the following, we present the variant of QCDCL implemented in
\depqbfnewversion that relies on a generalization of the Q-resolution calculus
(\qrescalc). We
illustrate the workflow of that variant in Fig.~\ref{fig:qcdcl:enhanced}.

In general, QCDCL is based on the successive generation of assignments that
guide the application of the inference rules of \qrescalc to derive
\emph{learned clauses} and \emph{cubes} from a
given input PCNF $\qclauset = \prefix. \clauset$. Learned cubes are dual to 
clauses. While learned clauses represent assignments that falsify the CNF
$\clauset$ of $\psi$, learned cubes represent assignments that satisfy
$\clauset$. 
The empty cube is derivable
from a PCNF $\qclauset$ by \qrescalc iff $\qclauset$ is
satisfiable. 
Based on our presentation of the rules of \qrescalc 
we illustrate the differences between traditional
QCDCL and the variant implemented in
\depqbfnewversion.

A QCDCL solver maintains a PCNF $\LCL = \prefix.\LCL'$ (PDNF $\LCU =
\prefix.\LCU'$) consisting of a CNF $\LCL'$ (DNF $\LCU'$) of learned clauses
(cubes).  The clauses in $\LCL$ are added conjunctively to $\qclauset$ to
obtain $\qclauset_{\LCL} = \prefix.(\clauset \wedge (\bigwedge_{C
\in \LCL'} C))$, and the cubes in $\LCU$ are added disjunctively to $\qclauset$
to obtain $\qclauset_{\LCU} = \prefix.(\clauset \vee (\bigvee_{C \in \LCU'}
C))$.  It holds that $\qclauset \satequiv \qclauset_{\LCL}$ and $\qclauset
\satequiv \qclauset_{\LCU}$.  Initially the current assignment $A$, the PCNF
$\LCL$, and PDNF $\LCU$ are empty.  We use the notation $C$, $C'$, and $C_L$
for both clauses and cubes.

During \emph{propagation}, the formulas $\qclauset_{\LCL}$ and
$\qclauset_{\LCU}$ are first simplified under the current assignment $A$ by
computing $\qclauset_{\LCL}[A]$ and $\qclauset_{\LCU}[A]$.  Then
\emph{universal} and \emph{existential reduction} is applied to
$\qclauset_{\LCL}[A]$ and to $\qclauset_{\LCU}[A]$ based on the following
inference rule.

\begin{definition}[Reduction~\cite{DBLP:journals/jair/GiunchigliaNT06,DBLP:journals/iandc/BuningKF95,DBLP:conf/tableaux/Letz02,DBLP:conf/iccad/ZhangM02}]Let $\qclauset = \prefix.\clauset$ be a PCNF.
\begin{align}\tag{$\mathit{red}$}
\AxiomC{$C \cup \{l\}$}
\UnaryInfC{C}
\label{rule_red}
\DisplayProof
\quad
\begin{minipage}{0.7625\textwidth}
(1) $C$ is a
clause, $\quant{\prefix}{l} = \forall$, \\
\hspace*{0.5cm} 
$l' <_\prefix l$ 
for all $l' \in C$ with $\quant{\prefix}{l'} = \exists$
or \\ 
(2) $C$ is a
cube, $\quant{\prefix}{l} = \exists$, \\
\hspace*{0.5cm} 
$l' <_\prefix l$ 
for all $l' \in C$ with $\quant{\prefix}{l'} = \forall$
\end{minipage} 
\end{align}
\end{definition}

Universal (existential) reduction of clauses (cubes) by rule~\ref{rule_red}
eliminates trailing universal (existential) literals from a
clause (cube)  with respect to the linear quantifier ordering in the prefix of the PCNF $\qclauset$.  We
write $\UR(C) = C'$ ($\ER(C) = C'$) to denote the clause (cube) $C'$ resulting
from clause (cube) $C$ by fully reducing universal (existential) literals.

Let $\qclauset_{\LCL}'$ and $\qclauset_{\LCU}'$ denote the formulas obtained
by applying universal (existential) reduction to all the clauses (cubes) in
$\qclauset_{\LCL}[A]$ ($\qclauset_{\LCU}[A]$) until saturation.  New
assignments are generated by \emph{unit literal detection} with respect to
$\qclauset_{\LCL}'$ and $\qclauset_{\LCU}'$.  If a PCNF
(PDNF) $\qclauset$ contains a \emph{unit clause (cube)} $C = (l)$, where
$\quant{\prefix}{l} = \exists$ ($\quant{\prefix}{l} = \forall$), then literal
$l$ is \emph{unit} and $\qclauset \satequiv \qclauset[A']$ where $A' = \{l\}$
($A' = \{\bar l\}$). 
Assignment $A$ is extended by assignments $A'$ derived from unit clauses
(cubes) in $\qclauset_{\LCL}'$ ($\qclauset_{\LCU}'$).
For every unit clause (cube)
$C' \in \qclauset_{\LCL}'$ ($C' \in \qclauset_{\LCU}'$) with $C' = (l)$, 
the corresponding assignment $A' := \{l\}$ ($A' := \{\bar l\}$) is recorded.

After propagation, in \emph{conflict/solution detection} it is checked whether
$\qclauset_{\LCL}'$ is unsatisfiable or whether $\qclauset_{\LCU}'$ is
satisfiable (only one of the two cases can occur). To this end, \emph{incomplete} methods are applied. In traditional
QCDCL, for example, it is \emph{syntactically} checked if the current
assignment $A$ is falsifying or satisfying, i.e., whether $\qclauset_{\LCL}'$ contains the empty clause (i.e., $\qclauset_{\LCL}' =
\bot$) or whether $\qclauset_{\LCU}'$ contains the empty cube (i.e., $\qclauset_{\LCU}' = \top$).  In \depqbfnewversion, we extend these
incomplete syntactic checks to incomplete \emph{semantic} checks based on \emph{arbitrary} QBF decision procedures
(proof systems) 
that are applied to $\qclauset_{\LCL}'$ and $\qclauset_{\LCU}'$ in a resource bounded \nolinebreak way. 

 If neither $\qclauset_{\LCL}'$ is found unsatisfiable nor $\qclauset_{\LCU}'$
 is found satisfiable by the incomplete satisfiability checks, then in \emph{decision
  making} $A$ is extended by heuristically assigning some \emph{decision variable} $x$ from the leftmost
quantifier block of $\qclauset[A]$ ($A := A \cup \{l\}$ where
$\lit{l} = x$), 
and propagation continues. Assignments by decision
making must follow the prefix ordering of $\qclauset$, in contrast to assignments by
propagation (unit literals), which results in assignments of the following kind. 
\begin{definition}[QCDCL assignment~\cite{DBLP:conf/lpar/LonsingBBES15}]\label{def:qcdcl:assignment}
Assignments generated by decision making and
propagation in QCDCL are called \emph{QCDCL assignments}.
\end{definition}

If $\qclauset_{\LCL}'$ ($\qclauset_{\LCU}'$) is found unsatisfiable
(satisfiable) in conflict/solution detection then a
learned clause
(cube) is derived using \qrescalc depending on the incomplete
satisfiability checks. 
In traditional QCDCL, conflict/solution detection relies only on 
falsifying or satisfying assignments. If $\qclauset_{\LCL}' = \bot$ then
$\qclauset_{\LCL}'$ contains an empty clause $C' = \emptyset$ such that there
is a clause $C \in \qclauset_{\LCL}$ with $C' = \UR(C[A])$.  Clause $C$ is the
\emph{falsified clause} with respect to assignment $A$. If $C$
appears in the given PCNF $\qclauset$ then in traditional \qrescalc it is
derived trivially by the following axiom.

\begin{definition}[Clause Axiom~\cite{DBLP:journals/jair/GiunchigliaNT06,DBLP:journals/iandc/BuningKF95,DBLP:conf/tableaux/Letz02,DBLP:conf/iccad/ZhangM02}] \label{def_trad_clause_axiom}
Let $\qclauset = \prefix.\clauset$ be a PCNF. 
\begin{align}\tag{$\operatorname{\emph{cl-init}}$}
\AxiomC{\phantom{A}}
\UnaryInfC{C}
\label{rule_cl_init}
\DisplayProof
\quad
\begin{minipage}{0.79\textwidth}
$C$ is a clause and $C \in \clauset$
\end{minipage} 
\end{align}
\end{definition}

If $\qclauset_{\LCL}' \not = \bot$ but $\qclauset_{\LCU}' = \top$ then either
(1) $\qclauset_{\LCU}'$ contains an empty learned cube $C' = \emptyset$ 
such that there is a cube $C \in \qclauset_{\LCU}$ with
$C' = \ER(C[A])$, or (2) $A$ is a satisfying assignment that satisfies all clauses in
$\qclauset_{\LCU}'$. For case (2), a cube $C$ is derived by the
following axiom of traditional \qrescalc 
(in either case (1) or (2) cube $C$ is the \emph{satisfied
cube} with respect to $A$).

\begin{definition}[Cube Axiom~\cite{DBLP:journals/jair/GiunchigliaNT06,DBLP:journals/iandc/BuningKF95,DBLP:conf/tableaux/Letz02,DBLP:conf/iccad/ZhangM02}] \label{def_trad_cube_axiom}
Let $\qclauset = \prefix.\clauset$ be a PCNF. 
\begin{align}\tag{$\operatorname{\emph{cu-init}}$}
\AxiomC{\phantom{A}}
\UnaryInfC{C}
\label{rule_cu_init}
\DisplayProof
\quad
\begin{minipage}{0.79\textwidth}
$A$ is an assignment,  
$\qclauset[A] = \top$, 
and $C = (\bigwedge_{l \in A} l)$ is a cube
\end{minipage} 
\end{align}
\end{definition}

\depqbfnewversion supports the application of \emph{arbitrary} (incomplete)
QBF decision procedures (proof systems) in conflict/solution detection and thus generalizes
the syntactic checks for falsifying and satisfying assignments  in
traditional QCDCL. 

To check the satisfiability of $\qclauset_{\LCU}'$, in \depqbfnewversion we apply a
dynamic variant of \emph{blocked clause elimination
(QBCE)}~\cite{DBLP:conf/lpar/LonsingBBES15}. This approach was
introduced in version~5.0 of \depqbf.  
QBCE has been presented as a
preprocessing technique to eliminate redundant \emph{blocked clauses}
\cite{DBLP:journals/jair/HeuleJLSB15,DBLP:journals/dam/Kullmann99} from a
PCNF. If all clauses in $\qclauset_{\LCU}'$ are satisfied under $A$ or
identified as blocked, then $\qclauset_{\LCU}'$ is determined
satisfiable. In our implementation applications of QBCE are tightly integrated in the propagation
phase via efficient data structures. Clauses that are blocked are
temporarily considered as removed from the formula. Hence such clauses
cannot be used to detect unit clauses or empty clauses during
propagation.

In addition to dynamic QBCE, we implemented incomplete QBF satisfiability
checks based on propositional abstractions of $\qclauset_{\LCL}'$ and
$\qclauset_{\LCU}'$~\cite{DBLP:conf/sat/LonsingES16}, which are solved using
an integrated SAT solver. These abstractions are constructed by treating
universally quantified literals in the given PCNF $\qclauset$ in a special
way. Propositional abstractions and SAT solving leverage the benefits of
techniques like \emph{trivial truth} and \emph{trivial
falsity} presented already in early search-based QBF
solvers~\cite{DBLP:conf/aaai/CadoliGS98}. Additionally, the power of
\emph{QU-resolution}~\cite{DBLP:conf/cp/Gelder12}, which is exponentially
stronger than Q-resolution~\cite{DBLP:journals/iandc/BuningKF95} but has not
been applied systematically in QCDCL, is harnessed to a certain extent (cf.~Example~3 in~\cite{DBLP:conf/sat/LonsingES16}).

As a simple way of applying a QBF decision procedure that is incomplete by its
nature we integrated the preprocessor
\bloqqer~\cite{DBLP:journals/jair/HeuleJLSB15} in \depqbfnewversion. 
Preprocessing aims at simplifying a formula within a restricted
amount of time but might already solve certain formulas 
(cf.~\cite{Lonsing201692}). Among several techniques, \bloqqer applies bounded
expansion of universally quantified
variables~\cite{DBLP:conf/sat/Biere04a,DBLP:conf/sat/BubeckB07}. Hence by
integrating \bloqqer in QCDCL we in fact integrate expansion, a QBF proof
system that is orthogonal to Q-resolution~\cite{beyersdorff_et_al:LIPIcs:2015:4905,DBLP:journals/tcs/JanotaM15}. 
Due to usability issues, in the follow-up release version 6.02 of \depqbf we replaced \bloqqer by the
expansion based QBF solver
\nenofex,\footnote{\url{https://github.com/lonsing/nenofex}} which is applied
in a resource bounded way.

If $\qclauset_{\LCL}'$ ($\qclauset_{\LCU}'$) is found unsatisfiable
(satisfiable) by an incomplete decision procedure but, unlike above, $A$ is neither falsifying nor satisfying, then a clause
(cube) is derived by the following \emph{generalized} axioms of
\qrescalc. These axioms are added to \qrescalc and applied in addition to the traditional
axioms~\ref{rule_cl_init} and~\ref{rule_cu_init}.

\begin{definition}[Generalized Axioms~\cite{DBLP:conf/sat/LonsingES16}] \label{def_gen_axioms}
Let $\qclauset = \prefix.\clauset$ be a PCNF. 
\begin{align}\tag{$\operatorname{\emph{gen-cl-init}}$}
\AxiomC{\phantom{A}}
\UnaryInfC{C}
\label{rule_gen_cl_init}
\DisplayProof
\quad
\begin{minipage}{0.7\textwidth}
$A$ is a QCDCL assignment, 
$\qclauset[A]$ is unsatisfiable, \\
and $C = (\bigvee_{l \in A} \bar l)$ is a clause
\end{minipage} 
\end{align}

\begin{align}\tag{$\operatorname{\emph{gen-cu-init}}$}
\AxiomC{\phantom{A}}
\UnaryInfC{C}
\label{rule_gen_cu_init}
\DisplayProof
\quad
\begin{minipage}{0.7\textwidth}
$A$ is a QCDCL assignment, 
$\qclauset[A]$ is satisfiable, \\
and $C = (\bigwedge_{l \in A} l)$ is a cube
\end{minipage} 
\end{align}
\end{definition}
Note that the generalized axioms allow to derive clauses and cubes that cannot be
derived by the traditional axioms~\ref{rule_cl_init}
and~\ref{rule_cu_init} in general. This is due to the application of
arbitrary QBF decision procedures (proof systems) for satisfiability checking
in conflict/solution detection or in the side conditions of the axioms, respectively. In the side conditions the satisfiability of the PCNF $\qclauset[A]$ is checked, in contrast to formulas $\qclauset_{\LCL}'$ and $\qclauset_{\LCU}'$ as in conflict/solution detection. This is possible since $\qclauset_{\LCL}' \satequiv \qclauset[A]$ and $\qclauset_{\LCU}' \satequiv \qclauset[A]$.
The clause (cube) $C$ derived by applying the generalized clause
axiom~\ref{rule_gen_cl_init} (\ref{rule_gen_cu_init}) is the \emph{falsified
clause (satisfied cube)} with respect to $A$.

During \emph{clause (cube) learning}, a new \emph{learned clause (cube)}
$C_{L}$ is derived by \qrescalc. The falsified clause (satisfied cube) $C$ is
the start clause (cube) of a derivation of $C_{L}$. Given $A$, 
clauses (cubes) which became unit during propagation are systematically resolved
based on the following \emph{Q-resolution} rule.
\begin{definition}[Q-Resolution~\cite{DBLP:journals/iandc/BuningKF95}] Let $\qclauset = \prefix.\clauset$ be a PCNF.  
\begin{align}\tag{$\mathit{res}$}
\AxiomC{$C_1 \cup \{p\}$}
\AxiomC{$C_2 \cup \{\bar p\}$}
\BinaryInfC{$C_1 \cup C_2$}
\label{rule_res}
\DisplayProof
\quad
\begin{minipage}{0.57\textwidth}
For all $x \in \prefix \colon \{x, \bar x\} \not \subseteq (C_1 \cup C_2)$, 
\\
$\bar p \not \in C_1$, $p
\not \in C_2$, and \emph{either} \\\hspace*{0.25cm} (1) $C_1$, $C_2$ are
clauses and $\quant{\prefix}{p} = \exists$ \emph{or} \\\hspace*{0.25cm}  (2) $C_1$, $C_2$ are cubes and $\quant{\prefix}{p} = \forall$
\end{minipage} 
\end{align}
\end{definition}
Rule~\ref{rule_res} does not allow the resolvent $(C_1 \cup C_2)$ to be a tautological clause (contradictory cube) and requires existential (universal) variables as pivots $p$.  In
general, learning produces a nonempty clause (cube) $C_{L} \not = \emptyset$,
which is added to the PCNF $\LCL$ (PDNF $\LCU$) of learned clauses (cubes),
and hence  also  to  $\qclauset_{\LCL}$ ($\qclauset_{\LCU}$).

In \emph{backtracking}, a certain subassignment $A' \subset A$ is
retracted such that $C_{L}$ becomes unit in propagation. $C_{L}$ is called an
\emph{asserting} clause (cube)~\cite{DBLP:journals/jair/GiunchigliaNT06}. 
Clauses (cubes) derived by rules~\ref{rule_cl_init}
and~\ref{rule_gen_cl_init} (\ref{rule_cu_init}
and~\ref{rule_gen_cu_init}) are used in exactly the same way in
learning to produce asserting clauses (cubes).

QCDCL terminates (``UNSAT'' or ``SAT'' in
Fig.~\ref{fig:qcdcl:enhanced}) by deriving the empty learned clause (cube)
$C_{L} = \emptyset$. A \emph{clause (cube) resolution proof} of the
unsatisfiability (satisfiability) of $\qclauset$ can be
obtained from the derivations of the learned clauses (cubes) up to
the empty clause (cube).

By applying the generalized axioms using a complete QBF decision
procedure, the empty assignment $A$, and an unlimited amount of time, 
the empty clause (cube) can  
be derived right away from any given unsatisfiable (satisfiable) PCNF
$\qclauset$. 
\emph{In practice} it is crucial to apply incomplete polynomial time procedures  
to limit the time spent on the satisfiability checks.  
However,  
the costs of frequent checks may outweigh the benefits. Hence
in \depqbfnewversion, satisfiability checks for applications of
the generalized axioms are dynamically disabled if they turn out to be too
costly, and the traditional axioms are used instead. We refer to
related work for implementation details~\cite{DBLP:conf/lpar/LonsingBBES15,DBLP:conf/sat/LonsingES16}.


\section{Features of \depqbf} \label{sec_features}

We briefly summarize the general features of \depqbf that have been
incorporated since its initial version~0.1~\cite{DBLP:journals/jsat/LonsingB10,DBLP:conf/sat/LonsingB10}. Most features were described
in related publications. Additionally, we comment on the
compatibility of the features with the implementation of \qrescalc
with generalized axioms (Fig.~\ref{fig:qcdcl:enhanced}) in
\depqbfnewversion.

\paragraph{\textbf{Dependency Schemes.}} 
Since the initial version~0.1, \depqbf has been equipped with the
\emph{standard dependency scheme}~\cite{DBLP:journals/jar/SamerS09} to
relax the linear quantifier ordering in the prefix of a given PCNF
$\qclauset$. In general, \emph{dependency
  schemes} are used to compute
\emph{dependency relations} $D$, which are binary relations over the
set of variables in $\qclauset$. If $(x,y) \not \in D$ for two
variables $x$ and $y$ then the ordering of $x$ and $y$ in $\qclauset$
can safely be swapped. Otherwise, if $(x,y) \in D$ then $y$ is
considered to depend on $x$. The integration of dependency schemes in
QCDCL results in the following reduction rule, which is added to
\qrescalc and implemented in \depqbf.

\begin{definition}[Dependency-Aware Reduction~\cite{DBLP:conf/sat/LonsingB10}] Let $\qclauset = \prefix.\clauset$ be a PCNF and $D$ be a dependency relation computed using a dependency scheme.
\begin{align}\tag{$\operatorname{\emph{dep-red}}$}
\AxiomC{$C \cup \{l\}$}
\UnaryInfC{C}
\label{rule_dep_red}
\DisplayProof
\quad
\begin{minipage}{0.70\textwidth}
(1) $C$ is a
clause, $\quant{\prefix}{l} = \forall$, \\
\hspace*{0.5cm} 
$(l,l') \not \in D$  
for all $l' \in C$ with $\quant{\prefix}{l'} = \exists$
or \\ 
(2) $C$ is a
cube, $\quant{\prefix}{l} = \exists$, \\
\hspace*{0.5cm} 
$(l,l') \not \in D$  
for all $l' \in C$ with $\quant{\prefix}{l'} = \forall$
\end{minipage} 
\end{align}
\end{definition}
Rule~\ref{rule_dep_red} generalizes the traditional reduction
rule~\ref{rule_red} by the use of dependency relation instead of the
linear ordering of variables ($\leq_\prefix$) in the prefix of PCNF
$\qclauset$. This way, it might be possible to reduce literals by
rule~\ref{rule_dep_red} which cannot be reduced by
rule~\ref{rule_red}. The soundness of \qrescalc with
rule~\ref{rule_dep_red} has been proved for a dependency relation that
is even more general (and thus allows for additional reductions) than
the one implemented in
\depqbf~\cite{DBLP:conf/sat/SlivovskyS12,DBLP:journals/tcs/SlivovskyS16,DBLP:conf/cp/Gelder11}. The
generalized axioms~\ref{rule_gen_cl_init} and~\ref{rule_gen_cu_init}
of \qrescalc implemented in \depqbfnewversion are naturally compatible
with rule~\ref{rule_dep_red}. Additionally, dependency schemes enable a
relaxed variant of QCDCL assignments (Definition~\ref{def:qcdcl:assignment})
based on the respective dependency relation rather than the prefix ordering of
a PCNF $\qclauset$.

\paragraph{\textbf{Long-Distance Resolution.}}
The Q-resolution
rule~\ref{rule_res}~\cite{DBLP:journals/iandc/BuningKF95} explicitly
disallows to generate clauses (cubes) that are tautological
(contradictory). This restriction is relaxed under certain
side conditions in \emph{long-distance (LD)
  Q-resolution}~\cite{DBLP:journals/fmsd/BalabanovJ12,DBLP:conf/iccad/ZhangM02,DBLP:conf/cp/ZhangM02}. LDQ-resolution
was first implemented in the QCDCL solver
\quaffle~\cite{DBLP:conf/iccad/ZhangM02} and was incorporated in
version~3.0 of \depqbf. Compared to \qrescalc with traditional
Q-resolution~\ref{rule_res}~\cite{DBLP:journals/iandc/BuningKF95},
\qrescalc with LDQ-resolution is exponentially more powerful in terms
of proof sizes~\cite{DBLP:conf/lpar/EglyLW13}. The generalized
axioms~\ref{rule_gen_cl_init} and~\ref{rule_gen_cu_init} implemented
in \depqbfnewversion are not only compatible with the LDQ-resolution
rule, but with \emph{any} variants of Q-resolution
(cf.~\cite{DBLP:conf/sat/BalabanovWJ14}). Recently, the soundness of
the combination of LDQ-resolution of clauses and dependency schemes in
\qrescalc has been
proved~\cite{DBLP:conf/cp/BeyersdorffB16,DBLP:conf/sat/PeitlSS16},
leaving the soundness of cube resolutions as an open
problem. Therefore, the combination of LDQ-resolution and dependency
schemes is not supported in \depqbfnewversion.

\paragraph{\textbf{Incremental Solving.}}
Since version~3.0, \depqbf has been equipped with an API in C and Java
for incremental solving of sequences $S := \langle \qclauset_0,
\ldots, \qclauset_n \rangle$ of syntactically related PCNFs
$\qclauset_i$~\cite{DBLP:conf/cp/LonsingE14,DBLP:conf/date/MarinMLB12}.
Incremental solving aims at reusing the clauses and cubes that were
learned when solving PCNF $\qclauset_i$ when it comes to solve the
PCNFs $\qclauset_j$ with $i < j$.  The API of \depqbf allows to modify
the PCNFs in $S$ by manipulating the quantifier prefix and adding or
removing sets of clauses in a stack-based way. Since version 4.0, it
is possible to add or remove sets of clauses
arbitrarily~\cite{DBLP:conf/sat/LonsingE15} 
and to extract \emph{unsatisfiable cores}, 
i.e., unsatisfiable subformulas of the PCNF $\qclauset_i$.   
At any time
when solving $\qclauset_i \in S$, the soundness property of QCDCL
(Section~\ref{sec:qcdcl}) that $\qclauset \satequiv \qclauset_{\LCL}$
and $\qclauset \satequiv \qclauset_{\LCU}$, where $\qclauset =
\qclauset_i$, must hold. To guarantee that property
when using the generalized axioms for incremental solving,
\depqbfnewversion currently only applies the generalized cube
axiom~\ref{rule_gen_cu_init} with dynamic QBCE used to check
satisfiability of $\qclauset_{\LCU}'$ in conflict/solution detection
(Fig.~\ref{fig:qcdcl:enhanced}). Although this configuration restricts the power of the
generalized axioms, it has improved incremental solving in the context
of QBF-based conformant planning~\cite{DBLP:journals/amai/EglyKLP17}. As it is unclear
how to use dependency schemes effectively in incremental solving,
their application is disabled in \depqbfnewversion.

\paragraph{\textbf{Generation of Proofs and Certificates.}}
QCDCL solvers can produce \emph{clause (cube) resolution proofs} of
the unsatisfiability (satisfiability) of PCNFs as a byproduct of
clause (cube) learning. Since
version~1.0~\cite{DBLP:conf/sat/NiemetzPLSB12}, \depqbf is capable of
producing proofs without employing dependency schemes by
rule~\ref{rule_dep_red}. Given a proof $P$ of a PCNF $\qclauset$, a
\emph{certificate} of $\qclauset$ can be extracted from $P$ by
inspecting the reduction steps by rule~\ref{rule_red} in
$P$~\cite{DBLP:journals/fmsd/BalabanovJ12}. A certificate of an
unsatisfiable (satisfiable) PCNF $\qclauset$ is given by a set of
Herbrand (Skolem) functions which represent the universal
(existential) variables in $\qclauset$. Applications of the
generalized axioms in QCDCL in general impose considerable
restrictions on the certificate extraction process. The
workflow~\cite{DBLP:journals/fmsd/BalabanovJ12} to extract a
certificate from $P$ was originally presented for traditional
\qrescalc proofs. If proof $P$ contains clauses (cubes) derived by
rule~\ref{rule_gen_cl_init} (rule~\ref{rule_gen_cu_init}), then $P$
may lack information needed to extract correct certificates. As a
result, \depqbfnewversion does not support cube resolution proof
generation combined with the generalized cube
axiom~\ref{rule_gen_cu_init}. However, it supports clause resolution
proof generation with the generalized clause
axiom~\ref{rule_gen_cl_init} provided that only propositional
abstractions and SAT solving are used for satisfiability checking in
the side condition of this axiom.

\paragraph{\textbf{Advanced Generation of Learned Clauses and Cubes.}}
The derivation of a single asserting clause (cube) starting from a falsified
clause (satisfied cube) as implemented in traditional
QCDCL~\cite{DBLP:journals/jair/GiunchigliaNT06,DBLP:conf/tableaux/Letz02,DBLP:conf/iccad/ZhangM02}
has an exponential worst case~\cite{DBLP:conf/cp/Gelder12}. Since version 2.0
\depqbf comes with an approach that avoids this exponential
case~\cite{DBLP:conf/sat/LonsingEG13} by a revised selection of clauses
(cubes) to be resolved in learning. This advanced approach is compatible with
all the techniques presented above.


\section{Experiments} \label{sec_experiments}

We compare variants of \depqbfnewversioncurrent, which is the latest
follow-up release of
\depqbfnewversion, to top performing solvers of 
QBFEVAL'16~\cite{DBLP:conf/sat/Pulina16}. As benchmarks we consider
all 825 instances from the PCNF track, both in original form
(Table~\ref{fig:exp:825:instances:noprepro}) and preprocessed by
\bloqqer version 37
(Table~\ref{fig:exp:825:instances:prepro}). 
We take preprocessing into account as it might have a positive impact
on certain solvers while a negative on
others (cf.~\cite{Lonsing201692,DBLP:journals/fuin/MarinNPTG16}). 
Experiments were run on an AMD Opteron
6238 processor (2.6 GHz) under 64-bit Ubuntu Linux 12.04 with time and
memory limits of 1800 seconds and seven GB. Exceeding the memory limit
is counted as a time out.\footnote{We refer to the appendix of this paper with
  additional experimental results.}

\begin{table}[t]
\caption{Solved instances (\emph{S}), solved unsatisfiable
  (\emph{$\bot$}) and satisfiable ones (\emph{$\top$}), uniquely
  solved ones among all solvers (\emph{U}), and total wall clock time including time
  outs on 825 PCNFs from QBFEVAL'16 without
  (\ref{fig:exp:825:instances:noprepro}) and with preprocessing by \bloqqer (\ref{fig:exp:825:instances:prepro}).}
\addtocounter{table}{-1}
\begin{minipage}[b]{0.48\textwidth}
\begin{center}
\subfloat[Original instances.]{
{\setlength\tabcolsep{0.15cm}
\begin{tabular}{l@{\quad}c@{\quad}c@{\quad}c@{\quad}r@{\quad}c}
\hline
\emph{Solver} & \emph{S} & \emph{$\bot$} & \emph{$\top$} & \emph{U} & \emph{Time} \\
\hline
\aigsolve & 603 & 301 & 302 & 34 & 440K \\
\ghostqcegar & 593 & 292 & 301 & 7 & 457K \\
\qstsbreaksym & 578 & 294 & 284 & 3 & 469K \\
\depqbfbat & 458 & 255 & 203 & 0 & 682K \\
\depqbfdepmansimpleldq & 458 & 257 & 201 & 2 & 686K \\
\depqbfdepmansimple & 456 & 255 & 201 & 0 & 686K \\
\depqbfnoclauseaxioms & 448 & 246 & 202 & 0 & 703K \\
\depqbfnoqbce & 397 & 228 & 169 & 0 & 788K \\
\depqbfnocubeaxioms & 393 & 229 & 164 & 0 & 796K \\
\depqbfnoaxioms & 383 & 221 & 162 & 0 & 814K \\
\caqe & 378 & 202 & 176 & 9 & 831K \\
\qesto & 369 & 210 & 159 & 0 & 864K \\
\rareqs & 341 & 211 & 130 & 2 & 891K \\
\hline
\end{tabular}
}
\label{fig:exp:825:instances:noprepro}
}
\end{center}
\end{minipage}
\hfill
\begin{minipage}[b]{0.48\textwidth}
\begin{center}
\subfloat[Preprocessed by \bloqqer.]{
{\setlength\tabcolsep{0.15cm}
\begin{tabular}{l@{\quad}c@{\quad}c@{\quad}c@{\quad}r@{\quad}r}
\hline
\emph{Solver} & \emph{S} & \emph{$\bot$} & \emph{$\top$} & \emph{U} & \emph{Time} \\
\hline
\qstsbreaksym & 633 & 330 & 303 & 11 & 365K \\
\rareqs & 633 & 334 & 299 & 8 & 375K \\
\qesto & 620 & 321 & 299 & 0 & 395K \\
\depqbfnoclauseaxioms & 601 & 303 & 298 & 0 & 428K \\
\depqbfbat & 601 & 301 & 300 & 0 & 429K \\
\depqbfdepmansimpleldq & 598 & 300 & 298 & 2 & 437K \\
\depqbfdepmansimple & 597 & 299 & 298 & 0 & 436K \\
\caqe & 596 & 301 & 295 & 4 & 451K \\
\depqbfnoaxioms & 593 & 296 & 297 & 0 & 444K \\
\depqbfnocubeaxioms & 591 & 297 & 294 & 0 & 455K \\
\depqbfnoqbce & 587 & 293 & 294 & 0 & 455K \\
\ghostqcegar & 570 & 282 & 288 & 0 & 485K \\
\aigsolve & 567 & 286 & 281 & 14 & 481K \\
\hline
\end{tabular}
}
\label{fig:exp:825:instances:prepro}
}
\end{center}
\end{minipage}
\label{fig:exp:825:instances}
\refstepcounter{table}
\end{table}

\begin{table}[t]
\caption{Related to Table~\ref{fig:exp:825:instances:noprepro}: solver
performance on 402 filtered original (\emph{not preprocessed}) instances partitioned into
261 instances with at most two (\ref{fig:exp:bloqqer37:classes:few:noprepro})
and 141 with three or more quantifier alternations
(\ref{fig:exp:bloqqer37:classes:many:noprepro}).}
\addtocounter{table}{-1}
\begin{minipage}[b]{0.48\textwidth}
\begin{center}
\subfloat[At most two quantifier alternations.]{
{\setlength\tabcolsep{0.15cm}
\begin{tabular}{l@{\quad}r@{\quad}r@{\quad}r@{\quad}r@{\quad}c}
\hline
\emph{Solver} & \multicolumn{1}{l}{\emph{S}} & \emph{$\bot$} & \multicolumn{1}{l}{\emph{$\top$}} & \emph{U} & \emph{Time} \\
\hline
\ghostqcegar & 176 & 75 & 101 & 5 & 171K \\ 
\aigsolve & 138 & 66 & 72 & 14 & 250K \\ 
\qstsbreaksym & 136 & 58 & 78 & 0 & 232K \\ 
\rareqs & 76 & 43 & 33 & 1 & 340K \\ 
\depqbfdepmansimple & 69 & 35 & 34 & 0 & 351K \\ 
\depqbfbat & 69 & 35 & 34 & 0 & 351K \\ 
\depqbfnoclauseaxioms & 68 & 35 & 33 & 0 & 354K \\ 
\depqbfdepmansimpleldq & 67 & 34 & 33 & 0 & 354K \\ 
\qesto & 66 & 37 & 29 & 0 & 359K \\ 
\depqbfnocubeaxioms & 53 & 24 & 29 & 0 & 378K \\ 
\depqbfnoaxioms & 52 & 24 & 28 & 0 & 378K \\ 
\depqbfnoqbce & 52 & 23 & 29 & 0 & 379K \\ 
\caqe & 43 & 17 & 26 & 3 & 397K \\ 
\hline
\end{tabular}
}
\label{fig:exp:bloqqer37:classes:few:noprepro}
}
\end{center}
\end{minipage}
\hfill
\begin{minipage}[b]{0.48\textwidth}
\begin{center}
\subfloat[Three or more quantifier alternations.]{
{\setlength\tabcolsep{0.15cm}
\begin{tabular}{l@{\quad}c@{\quad}c@{\quad}c@{\quad}r@{\quad}c}
\hline
\emph{Solver} & \emph{S} & \emph{$\bot$} & \emph{$\top$} & \emph{U} & \emph{Time} \\
\hline
\depqbfdepmansimpleldq & 81 & 50 & 31 & 2 & 120K \\
\depqbfbat & 79 & 47 & 32 & 0 & 119K \\  
\depqbfnoclauseaxioms & 79 & 47 & 32 & 0 & 120K \\ 
\depqbfdepmansimple & 78 & 47 & 31 & 0 & 123K \\ 
\qstsbreaksym & 72 & 44 & 28 & 3 & 132K \\ 
\depqbfnoqbce & 56 & 37 & 19 & 0 & 159K \\ 
\ghostqcegar & 56 & 31 & 25 & 2 & 160K \\ 
\depqbfnoaxioms & 55 & 36 & 19 & 0 & 159K \\ 
\depqbfnocubeaxioms & 55 & 36 & 19 & 0 & 159K \\ 
\aigsolve & 54 & 25 & 29 & 9 & 161K \\ 
\qesto & 49 & 33 & 16 & 0 & 179K \\ 
\caqe & 46 & 29 & 17 & 2 & 182K \\ 
\rareqs & 43 & 33 & 10 & 0 & 180K \\
\hline
\end{tabular}
}
\label{fig:exp:bloqqer37:classes:many:noprepro}
}
\end{center}
\end{minipage}
\label{fig:exp:bloqqer37:classes:noprepro}
\refstepcounter{table}
\end{table}

\begin{table}[t]
\caption{ Related to Table~\ref{fig:exp:825:instances:prepro}: solver
performance on 402 filtered and \emph{preprocessed} instances partitioned into 270
instances with at most two (\ref{fig:exp:bloqqer37:classes:few:prepro}) and
132 with three or more quantifier alternations
(\ref{fig:exp:bloqqer37:classes:many:prepro}).}
\addtocounter{table}{-1}
\begin{minipage}[b]{0.48\textwidth}
\begin{center}
\subfloat[At most two quantifier alternations.]{
{\setlength\tabcolsep{0.15cm}
\begin{tabular}{l@{\quad}c@{\quad}r@{\quad}r@{\quad}r@{\quad}c}
\hline
\emph{Solver} & \emph{S} & \emph{$\bot$} & \emph{$\top$} & \emph{U} & \emph{Time} \\
\hline
\rareqs & 157 & 79 & 78 & 8 & 227K \\ 
\qesto & 138 & 66 & 72 & 0 & 255K \\ 
\qstsbreaksym & 136 & 62 & 74 & 2 & 255K \\ 
\caqe & 118 & 49 & 69 & 2 & 298K \\ 
\ghostqcegar & 111 & 46 & 65 & 1 & 304K \\ 
\depqbfbat & 107 & 43 & 64 & 1 & 311K \\ 
\depqbfdepmansimple & 106 & 42 & 64 & 0 & 311K \\ 
\depqbfnoclauseaxioms & 105 & 43 & 62 & 0 & 312K \\ 
\depqbfnoaxioms & 105 & 41 & 64 & 0 & 313K \\ 
\depqbfdepmansimpleldq & 104 & 40 & 64 & 0 & 315K \\
\aigsolve & 102 & 49 & 53 & 7 & 313K \\  
\depqbfnoqbce & 102 & 39 & 63 & 0 & 322K \\ 
\depqbfnocubeaxioms & 102 & 40 & 62 & 0 & 323K \\ 
\hline
\end{tabular}
}
\label{fig:exp:bloqqer37:classes:few:prepro}
}
\end{center}
\end{minipage}
\hfill
\begin{minipage}[b]{0.48\textwidth}
\begin{center}
\subfloat[Three or more quantifier alternations.]{
{\setlength\tabcolsep{0.15cm}
\begin{tabular}{l@{\quad}c@{\quad}c@{\quad}c@{\quad}r@{\quad}r}
\hline
\emph{Solver} & \emph{S} & \emph{$\bot$} & \emph{$\top$} & \emph{U} & \emph{Time} \\
\hline
\depqbfnoclauseaxioms & 83 & 51 & 32 & 0 & 96K \\ 
\depqbfbat & 81 & 49 & 32 & 0 & 98K \\ 
\depqbfdepmansimpleldq & 81 & 51 & 30 & 2 & 102K \\ 
\depqbfdepmansimple & 78 & 48 & 30 & 0 & 105K \\ 
\depqbfnocubeaxioms & 76 & 48 & 28 & 0 & 112K \\ 
\qstsbreaksym & 75 & 50 & 25 & 1 & 107K \\ 
\depqbfnoaxioms & 75 & 46 & 29 & 0 & 112K \\ 
\depqbfnoqbce & 72 & 45 & 27 & 0 & 113K \\ 
\qesto & 69 & 45 & 24 & 0 & 120K \\ 
\caqe & 64 & 42 & 22 & 0 & 136K \\ 
\rareqs & 62 & 45 & 17 & 1 & 131K \\ 
\aigsolve & 51 & 27 & 24 & 6 & 151K \\ 
\ghostqcegar & 46 & 26 & 20 & 0 & 162K \\ 
\hline
\end{tabular}
}
\label{fig:exp:bloqqer37:classes:many:prepro}
}
\end{center}
\end{minipage}
\label{fig:exp:bloqqer37:classes:prepro}
\refstepcounter{table}
\end{table}

To assess the impact of
the generalized axioms \ref{rule_gen_cl_init} and \ref{rule_gen_cu_init} on the performance, we consider \depqbfnewversioncurrent using
both \ref{rule_gen_cl_init} and \ref{rule_gen_cu_init} (variant \depqbfbat in the
tables), without \ref{rule_gen_cl_init} (\depqbfnoclauseaxioms), without  
\ref{rule_gen_cu_init} (\depqbfnocubeaxioms), and using no generalized
axioms at all (\mbox{\depqbfnoaxioms}). 

On original instances (Table~\ref{fig:exp:825:instances:noprepro}),
\depqbfbat outperforms variants \depqbfnoclauseaxioms,
\depqbfnocubeaxioms, and \depqbfnoaxioms with restricted or without 
generalized axioms, respectively. Variant
\depqbfnoclauseaxioms without axiom \ref{rule_gen_cl_init} outperforms
variant \depqbfnocubeaxioms without \ref{rule_gen_cu_init}. We
attribute this effect to the use of dynamic
QBCE (among other techniques) for applications of the cube axiom
\ref{rule_gen_cu_init} in \depqbfnoclauseaxioms. 
Compared to \depqbfbat, disabling
only dynamic QBCE in variant \depqbfnoqbce severely impacts performance.  

On preprocessed instances (Table~\ref{fig:exp:825:instances:prepro}),
we make similar observations regarding the impact of the generalized
axioms like in Table~\ref{fig:exp:825:instances:noprepro}.  However,
variant \depqbfnoclauseaxioms without the clause axiom
\ref{rule_gen_cl_init} is on par with  
\depqbfbat. Preprocessing may blur the structure of an instance. We
conjecture that this blurring
hinders the success of the QBF decision procedures in
\depqbf, on which applications of the generalized axioms are based. 
In general the performance
difference between the variants of
\depqbf is smaller than on original instances. The rankings of the
solvers \rareqs~\cite{Janota20161}, \qesto~\cite{DBLP:conf/ijcai/JanotaM15}, and \caqe~\cite{DBLP:conf/fmcad/RabeT15} are
improved substantially by preprocessing, whereas those of \aigsolve~\cite{DBLP:conf/sat/SchollP16}
and \ghostqcegar~\cite{Janota20161,DBLP:conf/sat/KlieberSGC10} become worse. The best variant \depqbfnoclauseaxioms
in Table~\ref{fig:exp:825:instances:prepro} ranks fourth behind \qstsbreaksym~\cite{DBLP:conf/sat/0001JT16},
\rareqs, and
\qesto. However, the lag to the
solver ranked third is 19 instances compared to 120 instances
for the best variant \depqbfbat in
Table~\ref{fig:exp:825:instances:noprepro} that also ranks fourth.

To analyze the effects of preprocessing in more detail, we filtered the 825
PCNFs from QBFEVAL'16 by discarding 354 PCNFs that are already solved by
\bloqqer and 69 PCNFs where \bloqqer eliminated all universally quantified
variables, resulting in a set of 402 PCNFs. Further, we considered the 402
PCNFs in their original form and preprocessed by \bloqqer and partitioned them
into subsets containing PCNFs with at most two and with three or more
quantifier alternations. Such partitioning is motivated by a related
experimental study~\cite{DBLP:journals/corr/LonsingE17} where a large
diversity of solver performance was observed on instance classes defined by
alternations. Tables~\ref{fig:exp:bloqqer37:classes:noprepro}
and~\ref{fig:exp:bloqqer37:classes:prepro} show solver performance on these
subsets without and with preprocessing, respectively. Notably, variants of
\depqbf outperform the other solvers on the subsets with three or more
alternations, both without and with preprocessing
(Tables~\ref{fig:exp:bloqqer37:classes:many:noprepro}
and~\ref{fig:exp:bloqqer37:classes:many:prepro}). 

All variants of \depqbf reported above apply dependency-aware reduction by
rule \ref{rule_dep_red}.  Variant \depqbfdepmansimple is the same as
\depqbfbat (including generalized axioms) but uses the traditional reduction
rule~\ref{rule_red} based on the linear quantifier ordering of PCNFs.  Variant
\depqbfbat outperforms \depqbfdepmansimple in all tables
except Table~\ref{fig:exp:bloqqer37:classes:few:noprepro}, where \depqbfdepmansimple
is on par, which illustrates the benefits of dependency schemes in
QCDCL. Variant \depqbfdepmansimpleldq differs from \depqbfdepmansimple in the
use of LDQ-resolution in learning instead of traditional Q-resolution by
rule~\ref{rule_res}. The results with LDQ-resolution are mixed, despite being
a stronger proof system than Q-resolution. Variant
\depqbfdepmansimpleldq outperforms \depqbfdepmansimple in all tables
except Tables~\ref{fig:exp:bloqqer37:classes:few:noprepro}
and~\ref{fig:exp:bloqqer37:classes:few:prepro}, i.e., on instances with at
most two quantifier alternations.


\section{Conclusion}

We presented the latest major release 
version~6.0 of the QCDCL solver \depqbf. \depqbfnewversion implements a
variant of QCDCL that is based on a generalization of the Q-resolution
calculus (\qrescalc). The generalization is achieved by equipping
\qrescalc with generalized clause and cube axioms to be used in 
clause and cube learning~\cite{DBLP:conf/sat/LonsingES16}. The generalized
axioms provide an extensible framework of 
interfaces for the integration of arbitrary QBF proof systems in
\qrescalc, and hence in QCDCL. The integration of proof systems
orthogonal to Q-resolution, such as variable expansion, enables QCDCL
to potentially produce proofs that are exponentially shorter than
proofs produced by traditional QCDCL. This way, the state of the art 
of QCDCL solving can be further advanced. A related open
problem is the inability of plain QCDCL to exploit the full power of
Q-resolution~\cite{DBLP:conf/sat/Janota16}.

The workflow of QCDCL with generalized axioms is not tailored towards
\depqbfnewversion but can be implemented in any QCDCL
solver. Furthermore, it is compatible with dependency
schemes~\cite{DBLP:journals/jar/SamerS09,DBLP:journals/tcs/SlivovskyS16}
and any Q-resolution variant~\cite{DBLP:conf/sat/BalabanovWJ14},
which offers potential for further improvements.

Experiments with variants of \depqbfnewversion showed considerable
performance gains due to the application of generalized axioms. However,
frequent applications are hindered by computationally expensive QBF
satisfiability checks in the side conditions of the axioms. To limit the
checking overhead, axiom applications must be carefully scheduled.
In this respect, there is room for improvements in fine tuning
\depqbfnewversion. Further, it may be beneficial 
to integrate the QBF decision procedures that are
applied to satisfiability checking more tightly in the QCDCL workflow, like
with dynamic blocked clause elimination (QBCE)~\cite{DBLP:conf/lpar/LonsingBBES15}.



\newpage

\begin{appendix}

\section{Additional Experimental Data}

\begin{table}[ht]
\caption{Related to Table~\ref{fig:exp:825:instances:noprepro} (no
  preprocessing): the 825 PCNFs are partitioned into 466
  PCNFs with at most two (\ref{fig:exp:classes:few}), and 359 PCNFs
  with three or more quantifier alternations
  (\ref{fig:exp:classes:many}).}
\addtocounter{table}{-1}
\begin{minipage}[b]{0.48\textwidth}
\begin{center}
\subfloat[At most two quantifier alternations.]{
{\setlength\tabcolsep{0.15cm}
\begin{tabular}{l@{\quad}c@{\quad}r@{\quad}r@{\quad}r@{\quad}r}
\hline
\emph{Solver} & \emph{S} & \multicolumn{1}{l}{\emph{$\bot$}} & \multicolumn{1}{l}{\emph{$\top$}} & \emph{U} & \emph{Time} \\
\hline
\aigsolve & 338 & 170 & 168 & 25 & 262K \\ 
\ghostqcegar & 336 & 160 & 176 & 5 & 257K \\ 
\qstsbreaksym & 292 & 143 & 149 & 0 & 325K \\ 
\depqbfdepmansimple & 233 & 126 & 107 & 0 & 429K \\ 
\depqbfbat & 233 & 126 & 107 & 0 & 429K \\ 
\depqbfdepmansimpleldq & 231 & 125 & 106 & 0 & 431K \\ 
\rareqs & 231 & 140 & 91 & 2 & 434K \\ 
\depqbfnoclauseaxioms & 223 & 117 & 106 & 0 & 449K \\ 
\qesto & 213 & 126 & 87 & 0 & 474K \\ 
\depqbfnoqbce & 202 & 110 & 92 & 0 & 483K \\ 
\depqbfnocubeaxioms & 197 & 111 & 86 & 0 & 492K \\ 
\depqbfnoaxioms & 188 & 103 & 85 & 0 & 509K \\ 
\caqe & 174 & 91 & 83 & 4 & 535K \\ 
\hline
\end{tabular}
}
\label{fig:exp:classes:few}
}
\end{center}
\end{minipage}
\hfill
\begin{minipage}[b]{0.48\textwidth}
\begin{center}
\subfloat[Three or more quantifier alternations.]{
{\setlength\tabcolsep{0.15cm}
\begin{tabular}{l@{\quad}c@{\quad}r@{\quad}r@{\quad}r@{\quad}c}
\hline
\emph{Solver} & \emph{S} & \multicolumn{1}{l}{\emph{$\bot$}} & \multicolumn{1}{l}{\emph{$\top$}} & \emph{U} & \emph{Time} \\
\hline
\qstsbreaksym & 286 & 151 & 135 & 3 & 144K \\
\aigsolve & 265 & 131 & 134 & 9 & 177K \\ 
\ghostqcegar & 257 & 132 & 125 & 2 & 199K \\ 
\depqbfdepmansimpleldq & 227 & 132 & 95 & 2 & 254K \\ 
\depqbfbat & 225 & 129 & 96 & 0 & 253K \\ 
\depqbfnoclauseaxioms & 225 & 129 & 96 & 0 & 254K \\ 
\depqbfdepmansimple & 223 & 129 & 94 & 0 & 257K \\ 
\caqe & 204 & 111 & 93 & 5 & 295K \\ 
\depqbfnocubeaxioms & 196 & 118 & 78 & 0 & 304K \\ 
\depqbfnoaxioms & 195 & 118 & 77 & 0 & 304K \\ 
\depqbfnoqbce & 195 & 118 & 77 & 0 & 305K \\ 
\qesto & 156 & 84 & 72 & 0 & 390K \\ 
\rareqs & 110 & 71 & 39 & 0 & 456K \\ 
\hline
\end{tabular}
}
\label{fig:exp:classes:many}
}
\end{center}
\end{minipage}
\label{fig:exp:classes}
\refstepcounter{table}
\end{table}

\begin{table}[ht]
\caption{Related to Table~\ref{fig:exp:825:instances:prepro}
  (preprocessed by \bloqqer): the 825 PCNFs are partitioned into 693 PCNFs with at
  most two (\ref{fig:exp:bloqqer37:classes:few}), and 132 PCNFs with
  three or more quantifier alternations
  (\ref{fig:exp:bloqqer37:classes:many}).}
\addtocounter{table}{-1}
\begin{minipage}[b]{0.48\textwidth}
\begin{center}
\subfloat[At most two quantifier alternations.]{
{\setlength\tabcolsep{0.15cm}
\begin{tabular}{l@{\quad}c@{\quad}r@{\quad}r@{\quad}r@{\quad}r}
\hline
\emph{Solver} & \emph{S} & \multicolumn{1}{l}{\emph{$\bot$}} & \multicolumn{1}{l}{\emph{$\top$}} & \emph{U} & \emph{Time} \\
\hline
\rareqs & 571 & 289 & 282 & 16 & 243K \\ 
\qstsbreaksym & 558 & 280 & 278 & 2 & 258K \\ 
\qesto & 551 & 276 & 275 & 0 & 274K \\ 
\caqe & 532 & 259 & 273 & 2 & 315K \\ 
\ghostqcegar & 524 & 256 & 268 & 1 & 323K \\ 
\depqbfbat & 520 & 252 & 268 & 1 & 331K \\ 
\depqbfdepmansimple & 519 & 251 & 268 & 0 & 331K \\ 
\depqbfnoaxioms & 518 & 250 & 268 & 0 & 332K \\ 
\depqbfnoclauseaxioms & 518 & 252 & 266 & 0 & 332K \\ 
\depqbfdepmansimpleldq & 517 & 249 & 268 & 0 & 334K \\
\aigsolve & 516 & 259 & 257 & 7 & 329K \\  
\depqbfnoqbce & 515 & 248 & 267 & 0 & 342K \\ 
\depqbfnocubeaxioms & 515 & 249 & 266 & 0 & 342K \\ 
\hline
\end{tabular}
}
\label{fig:exp:bloqqer37:classes:few}
}
\end{center}
\end{minipage}
\hfill
\begin{minipage}[b]{0.48\textwidth}
\begin{center}
\subfloat[Three or more quantifier alternations.]{
{\setlength\tabcolsep{0.15cm}
\begin{tabular}{l@{\quad}c@{\quad}c@{\quad}c@{\quad}r@{\quad}r}
\hline
\emph{Solver} & \emph{S} & \emph{$\bot$} & \emph{$\top$} & \emph{U} & \emph{Time} \\
\hline
\depqbfnoclauseaxioms & 83 & 51 & 32 & 0 & 96K \\ 
\depqbfbat & 81 & 49 & 32 & 0 & 98K \\ 
\depqbfdepmansimpleldq & 81 & 51 & 30 & 2 & 102K \\ 
\depqbfdepmansimple & 78 & 48 & 30 & 0 & 105K \\ 
\depqbfnocubeaxioms & 76 & 48 & 28 & 0 & 112K \\ 
\qstsbreaksym & 75 & 50 & 25 & 1 & 107K \\ 
\depqbfnoaxioms & 75 & 46 & 29 & 0 & 112K \\ 
\depqbfnoqbce & 72 & 45 & 27 & 0 & 113K \\ 
\qesto & 69 & 45 & 24 & 0 & 120K \\ 
\caqe & 64 & 42 & 22 & 0 & 136K \\ 
\rareqs & 62 & 45 & 17 & 1 & 131K \\ 
\aigsolve & 51 & 27 & 24 & 6 & 151K \\
\ghostqcegar & 46 & 26 & 20 & 0 & 162K \\ 
\hline
\end{tabular}
}
\label{fig:exp:bloqqer37:classes:many}
}
\end{center}
\end{minipage}
\label{fig:exp:bloqqer37:classes}
\refstepcounter{table}
\end{table}

\end{appendix}

\end{document}